\documentstyle[12pt]{article}
\begin{document}
\baselineskip = 20pt


\begin{center}

{\bf A SIMPLE LINE ELEMENT TO THE DILATON GRAVITY  }
\footnote{Work supported in part by 
Funda\c c\~ao Universit\'aria Jos\'e Bonif\'acio, FUJB.}
\vspace*{1cm}

\end{center}

\begin{center}

{M. Alves$^{a}$}

\end{center}

\begin{center}

{Instituto de F\'\i sica \\ Universidade Federal do Rio de Janeiro \\

 Caixa Postal 68528  Rio de Janeiro 21970-970 Brazil}

\end{center}

\vskip 1cm

\vspace*{5mm}
\begin{center}
{\bf ABSTRACT}
\end{center}
\vspace*{2mm}

\noindent

The metric to the two-dimensional dilaton gravity can be writen in an alternative form, similar to the two-dimensional Schwarzschild metric, and allow us the identification of some quantitiies with those equivalent in the Schwarzschild solution. This new form, howewer, presents a non-physical singularity at the horizon in the same way that in the realistic four dimensional case. We show a procedure to eliminate this horizon singularity and, as an application, the resulting metric is used to obtain the associated Hawking temperature. We discuss also some differents between this metric and the Schwarzcchild one.

.

\bigskip
\vfill
\par
\bigskip
\bigskip

\par
\bigskip
\bigskip
\noindent $(a)${msalves@if.ufrj.br}

\pagebreak

\vspace*{0.5cm}

\noindent{1- INTRODUCTION }

\bigskip
In the heuristic picture often proposed to visualize the origin of the Hawking-Bekenstein effect[1], the radiation of the black-hole arises by the combination of the pair creation and tunneling processes. At the same time, the simplest way to describe these objects to wit, the Schwarzschild solution has an (unphysical) singularity at the horizon.
So, since we need to describe the tunneling as an across-horizon phenomena, it is necessary to choose coordinates that, unlike Schwarzschild ones, are not singular at the horizon. These  coordinates  are the well known Kruskal-Szekeres coordinates.

We have the same situation in the two-dimensional dilaton gravity in the Schwarzrschild form where there is a coordinate singularity that has to be avoided to calculate some quantities at the horizon. 
As in the realistic case above mentioned, we can put the metric in such a way that this singularity is removed. 
This is the  purpose of this note: to write out a metric to the dilaton gravity without this coordinate singularity. As an application, we calculate the associate Hawking temperature in a very simple way.

The article is organized as follows. We show the coordinate transformation for the 2d Schwarzschild case. Then, with a similar transformation, we obtain the desired form to the dilaton metric. We finish with the application of this line element to the calculation of the associated temperature of the black-hole. 

\bigskip
\bigskip
\bigskip
\noindent{2- THE NEW COORDINATES TO THE TWO-DIMENSIONAL GRAVITY  }

\bigskip
We start this section with the Schwarzschild solution where the angular parts have been discarded:

\begin{equation}
ds^{2} = g_{\mu\nu}dx^{\mu}dx^{\nu}= -\left(1-{2M\over r}\right)dt_{s}^{2} + \left(1-{2M\over r}\right)^{-1}dr^{2}
\end{equation}

\noindent where $t_{s}$ means Schwarzschild coordinates. As in four dimensions, there is a real singularity at $r=0$ and a non-physical one at $r=2M$, where the horizon is localized.

\noindent A particularly suitable expression is obtained by introducing a new time coordinate defined [2] as

\begin{equation}
t=t_{s}+2\sqrt{2Mr} + 2M ln {\sqrt{r} - \sqrt{2M}\over \sqrt{r} + \sqrt{2M}}
\end{equation}

The line element turns to be

\begin{equation}
ds^{2} = -(1-{2M\over r})dt^{2} + 2\sqrt{2M\over r}dtdr + dr^{2}
\end{equation}

\noindent that, by a simple inspection, shows us that there is no more singularity at $r=2M$.

\bigskip
Let us apply the same procedure to the CGHS model [3] to the bidimensional gravity, the so called dilaton gravity.
The line element of this model is given by

\begin{equation}
ds^{2} = \left( {M\over \lambda} - \lambda^{2}xy\right)dx dy
\end{equation}

\noindent where $x = t + r$\, and $y = t -r$\, are the light cone coordinates. We can see that the singularity (real) is at 
\begin{equation}
xy = {M\over \lambda^{3}}
\end{equation}

\noindent and the horizon at $y=0$.

The following coordinate transformation  [4]

\begin{equation}
1-{2M\over r} = \left( 1 + {M\over\lambda}e^{-2\lambda\sigma}\right)^{-1}
\end{equation}

\noindent with 

\begin{equation}
\lambda x = e^{\lambda\sigma^{+}} \,\,\,\,\,\,\, , \,\,\,
\lambda y = e^{\lambda\sigma^{-}} 
\end{equation}

\noindent and $\sigma^{\pm} = t \pm \sigma$ puts the metric in the "Schwarzschild type" form:

\begin{equation}
ds^{2} = -\left(1-{2M\over r}\right) dt^{2} + {1\over 4\lambda^{2}r^{2}}\left(1-{2M\over r}\right)^{-1} dr^{2}
\end{equation}

In this way, the horizon is mapped to $r=2M$, the curvature singularity is at $r=0$ and the asymptotic regions are described by $r=\infty$. Note that there is a new dependence in $r$.

With the metric in a form very close to that of the Schwarzschid one,
the non-physical singularity $r=2M$ arises and if we want to make use of the resemblance between (8) and (1), we must find
  a transformation that makes (8) look like  expression (3).

This transformation is given by

\begin{equation}
t=t_{s}+{1\over 2\lambda} ln {\sqrt{r} - \sqrt{2M}\over \sqrt{r} + \sqrt{2M}}
\end{equation}

\noindent which put the line element as
\begin{equation}
ds^{2} = -(1-{2M\over r})dt_{s}^{2} + {1\over \lambda r}\sqrt{2M\over r}dtdr + {1\over 4\lambda^{2}r^{2}}dr^{2}
\end{equation}

\noindent without the $r=2M$ singularity.

\bigskip
\bigskip
\bigskip
\noindent
 {  {4- THE HAWKING TEMPERATURE}}

\bigskip

The thermodynamic associated with the black holes is mostly based on the properties on the horizon. This is the case of the flux of outgoing particles from these objects. From [5] the total flux of outgoing particles is given by

\begin{equation}
\rho(w_{k})\sim \vert \beta_{kk^{'}}/\alpha_{kk^{'}}\vert^{2}
\end{equation}

\noindent where $\beta_{kk^{'}}$ and  $\alpha_{kk^{'}}$ are the Bogoliubov coefficients given by

\begin{equation}
\alpha_{kk'}=\int^{\infty}_{-\infty} dt\, e^{iw_{k}t}v_{k'}(t,r) \,\,\,\,\,\,\,\,\,
and  \,\,\,\,\,\,\,\,\,\
\beta_{kk'}=\int^{\infty}_{-\infty} dt\, e^{-iw_{k}t}v_{k'}(t,r)
\end{equation}

Where the modes $v_{kk'}$ must be non singular at the horizon (the coordinates in (10) have a fundamental role here).

It is convenient to write these modes in terms of the Kruskal-type coordinates 

\begin{equation}
v_{k'}=e^{ik^{'}U}
\end{equation}

with $U$, defined in terms of the (9), gives a time dependence 
\begin{equation}
v_{k'} \sim exp ( -ik'e^{-\lambda t})
\end{equation}

\noindent The presence of $\lambda $  assures the non singular behavior of the metric on the horizon. This factor corresponds to the  
$1/4m$ in the Schwarzschild case.

 Following the same steps as in [5], we find that
\begin{equation}
\vert \beta_{kk^{'}}/\alpha_{kk^{'}}\vert^{2} = e^{-(2\pi  /\lambda ) w_{k}}
\end{equation}

\noindent and using (15) in (11)

\begin{equation}
\rho(w_{k})\sim e^{-(2\pi /\lambda)w_{k}}
\end{equation}

\noindent giving rise the picture of  the black-hole  as a thermal body at the (Hawking) temperature given by
\begin{equation}
T_{H} = {\lambda\over 2\pi}
\end{equation}

\bigskip
\bigskip
\bigskip
\noindent
 {  4- CONCLUSION}

\bigskip
\bigskip

This note deals with  a  suitable form of the metric to the dilaton gravity, obtained after two steps: the original metric is transformed  into a Schwarzschild-type, where is possible to identify in a simple way some of the relevant parameters as well as its asymptotic behavior. 
With another transformation we have eliminated the non physical singularity at the horizon.
This second transformation was already applied to the bidimensional Schwarzschild case and resembles that used in the definition of the Kruskall coordinates.

After that, we use the new expression to obtain the associated Hawking temperature. This new metric allows us to redefine coordinates in the same procedure used to the Schwarzschild four dimensional case.

\bigskip
\bigskip
\bigskip

 {  {REFERENCES}}

\noindent 1- S. W. Hawking, Commun.Math. Phys. 43, 199 (1975).

\noindent 2- P. Kraus and F. Wilczek, Mod. Phys. Lett. A 40 (1994) 3713;
 M. K. Parik and F. Wilczek, Hep-th/9907001. 

\noindent 3- C. Callan, S. Giddings, J. Harvey and A. Strominger, Phys. Rev. D 45, R1005 (1992). For recent review with updated references see S.Nojiri and S.O.Odintsov, Hep-th/0009202.

\noindent 4- A. Ghosh. Hep-th/9604056

\noindent 5- E.Keski-Vakkuri and P.Krauss.Nucl.Phys.B 491 (1997)249.

\end{document}